# Trilobite "pelotons": Possible hydrodynamic drag effects between leading and following trilobites in trilobite queues

HUGH TRENCHARD[1,*], CARLTON BRETT[2] *and* MATJAŽ PERC[3,4]

[1]805 647 Michigan Street, Victoria BC V8V 1S9
[2]Geology Department, University of Cincinnati, Ohio 45221-0013
[3]Faculty of Natural Sciences and Mathematics, University of Maribor, Koroška cesta 160, SI-2000 Maribor, Slovenia
[4]CAMTP - Center for Applied Mathematics and Theoretical Physics, University of Maribor, Krekova 2, SI-2000 Maribor, Slovenia

*corresponding author: h.a.trenchard@gmail.com*

Energy saving mechanisms in nature allow following organisms to expend less energy than leaders. Queues, or ordered rows of individuals, may form when organisms exploit the available energy saving mechanism while travelling at near-maximal sustainable metabolic capacities; compact clusters form when group members travel well below maximal sustainable metabolic capacities. The group size range, given here as the ratio of the difference between the size of the largest and smallest group members, and the size of the largest member (as a per cent), has been hypothesized to correspond proportionately to the energy saving quantity because weaker, smaller, individuals sustain speeds of stronger, larger, individuals by exploiting the energy saving mechanism (as a per cent). During migration, small individuals outside this range may perish, or form sub-groups, or simply not participate in migratory behavior. We approximate drag forces for leading and following individuals in queues of the late Devonian (~370 Ma) trilobite *Trimerocephalus chopini.* Applying data from literature of *R. herculea*, a living crustacean, we approximate the hypothetical walking speed and maximal speeds for *T. chopini*. Findings reasonably support the hypothesis: among the population of fossilized queues of *T. chopini* reported by Kin & Błażejowski (2013), trilobite size range was 75% while the size range within queues, was 63%; this corresponds reasonably with drag reductions in following positions that permit ~61.5% energy saving for trilobites following others in optimal low-drag positions. We model collective trilobite behavior associated with hydrodynamic drafting.

## INTRODUCTION

The fossil record reveals that some trilobite species formed aggregations (Speyer & Brett 1985, Karim & Westrop 2002, Radwański et al. 2009). Queuing behavior, the tendency to form single-file rows of individuals, has been observed in Early Ordovician *Agerina* and *Ampyx* trilobites (Chatterton & Fortey 2008), the giant Middle Ordovician trilobites *Bathycheilus*, *Salterocoryphe* and/or *Retamaspis* (Gutiérrez-Marco et al. 2009), the blind Late Devonian phacopid *Trimerocephalus chopini* (Kin & Błażejowski 2013; Błażejowski et al. 2016), fossilized enigmatic Cambrian arthropods (Hou et al. 2008; Xian-Guang et al. 2009), and in living spiny lobsters (Bill & Herrnkind 1976). Fossilized "beaten" trackways of probable eurypterids indicate similar queueing behavior (Draganits et al. 1998, (their Fig. 6); Braddy 2001).

The focus of this study is on *T. chopini*, for which the highest number of trilobites in collected fossilized queues was 19 (Błażejowski et al. 2016). Most queues were composed of the largest individuals, while the smaller individuals were in short queues of two individuals (Błażejowski et al. 2016). Queues were organized in straight single-files, or slightly twisted, or arched, either separated with no contact, or with head-to-tail contact and overlap to varying degrees (Radwański et al. 2009; Błażejowski et al. 2016), as shown in Fig. 1C, D. Queues have also been described in zig-zag or wavy patterns (Gutiérrez-Marco et al. 2009). Of course, such queuing behavior is almost certainly under-represented in the fossil record as very special conditions of synchronous mortality and rapid burial without seafloor disturbance are required to preserve such rows.

In addition to queue formations, aggregations of trilobite fossils appear in more compact non-linear "clusters", which Speyer & Brett (1985) defined as "a group of three or more trilobites along a single bedding plane in which adjacent individuals are no more than two centimeters from one another" (p. 90). The authors reported clusters comprising from three individuals to over 200, in which individuals were randomly orientated. Speyer & Brett (1985) proposed that fossilized cluster assemblages revealed synchronous moulting and possibly mating behavior. Because of random orientations and evidence of moulting within clusters, these groupings probably do not record active migration, although aggregation itself implies some, at least local, migration. However, other trilobite clusters with non-random directions of cephala could represent a distinct type of migratory behavior.

Migration may be defined as "the directional movement of individuals of one species between distinct locations. The timescale on which these movement cycles occur can span hours, days, months or years, or it can be multigenerational" (Liedvogel et al. 2011, p. 562, Box 1). Dingle & Drake (2007) identify migration as one category of animal collective movement, which "can be divided into those that occur within the home range and those that take the individual more or less permanently beyond it" (p.116). Trilobite queuing behavior may have occurred both during short range commutes within the home range to and from spatially separated resource patches or roost sites, or during long distance migration permanently away from the initial home range.

With respect to single-file formation as a mode of collective locomotion, extant spiny lobsters form queues (Bill & Herrnkind 1976) when travelling ocean floors. Bill & Herrnkind (1976) demonstrated that spiny lobsters obtain ~65% reductions in energy expenditure while travelling along ocean beds in single-file formations.

Spiny lobster queue formations are unique among benthic crustaceans (Herrnkind et al. 1973), but single-file travelling formations have been observed among other arthropods, including ants (e.g. Hansen & Klotz 2005, p. 135; Wilson 1959), juvenile spiders (Reichling 2000), whirligig beetles (Heinrich & Vogt 1980), weevil larvae (Fitzgerald et al. 2004), and a variety of caterpillars (Steinbauer 2009). Among these, ant single file formations have been modelled and studied in terms of energy optimization (Chadhuri & Nagar 2015), but we found no reports quantifying the energy savings obtained by such formations.

Amid the fossil record, there is evidence of high density (~150 per m$^2$) trackways of unidirectional migration among Late Jurassic isopods from the Crayssac Lagerstätte, of Germany, thought to have been imprinted in soft-to-firm mud of a temporarily emerged tidal flat (Gaillard et al. 2005, their fig. 11B). Given their high density and apparent near proximity to each other during locomotion, the trackways suggest compact collective configurations like those observed among bicycle pelotons, as shown in Fig.1B, although it is difficult to interpret such isopod collective motion as involving hydrodynamic drafting.

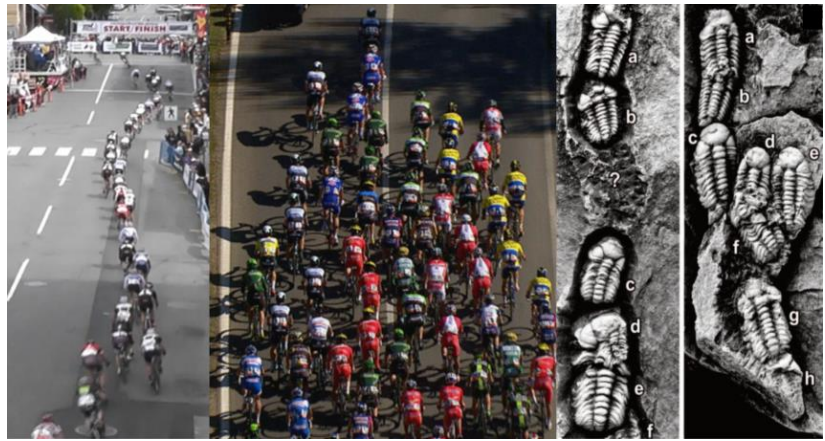

**Figure 1A.** Cyclists in single-file travel at near maximal sustainable outputs; **B.** Cyclists in unidirectional compact formation travel at below maximal sustainable outputs (from Trenchard & Perc, 2016, figs. 5B and 13A, with permission from the publisher); **C,** the trilobite *Trimerocephalus chopini* in single-file queue suggests trilobites travelling at speeds approaching maximal sustainable outputs (Radwański et al. 2009, pl. 4; Creative Commons Attribution License (CC BY)); **D**. formation somewhat more compact than in image **C**, which may show a transition state in which leaders are decelerating and being passed laterally by followers, producing an increasingly compact formation; or an acceleration by leaders producing increased queue stretching.

Bicycle peloton (group of cyclists) formations are generated largely by the energy saving mechanism of aerodynamic drafting. Drafting, either aerodynamically or hydro-dynamically, occurs when bodies, following in the wake of leading bodies, encounter reduced drag (Hoerner 1965). For cyclists, riding in zones of reduced air drag permits weaker cyclists, who may otherwise be incapable of such speeds in isolation, to sustain the speeds of stronger cyclists (Olds 1998; Trenchard et al. 2014, 2015).

Although driven in part by human-based competitive strategy, pelotons exhibit self-organized collective behaviors that emerge largely as a function of the metabolic outputs of the individuals within the group, and the power output reductions afforded by drafting (Trenchard et al. 2014). As cyclists approach their maximal sustainable capacities, formations stretch into single-file lines (queues); below a certain output threshold, single-file lines tend to collapse into compact unidirectional formations, as shown in Fig. 1A, B (Trenchard et al. 2014, 2015; Trenchard 2015).

Similar energy saving mechanisms involving hydrodynamic media have been demonstrated for a broad range of organism body sizes, which mechanisms facilitate generally unidirectional aggregations. For example, Grey mullet fish (*Liza aurata*) with body-lengths of ~12 cm obtain an energy saving of up 19%, where optimal energy saving positions are at angles to nearby fish (Marras et al. 2015). Mysid shrimp (*Paramesodopsis rufa*) with body-lengths 0.9 –1.3 cm, generate flow wakes and reduced energy requirements for those following in wakes (Ritz 2000). Wakes of similar-sized krill are shown also to project at angles to body orientation (Yen et al. 2000). At still smaller scales, drafting dynamics have been demonstrated for spherical inanimate particles in liquid, of diameters 0.1 – 0.2 cm, following directly behind leading particles (Wang et al. 2014). For a review of energy reduction mechanisms in a variety of biological and non-biological circumstances, exhibited across the entire range of animal sizes from bacteria to whales, see Trenchard & Perc (2016).

Hydrodynamic drafting appears to be a physical mechanism not yet considered in the trilobite literature as a driving mechanism of aggregative trilobite behavior, even though it has been demonstrated for modern arthropods travelling in queues (Bill & Herrnkind 1976). Here we consider the presence of hydrodynamic drag, drafting opportunities, and energetic differentials among trilobites in leading and following positions within queues.

We consider the behavioral consequences of these differentials, and their effects on the relative sizes of individual trilobites. In this context, certain scaling rules are applicable: except for birds and very large animals, speeds tend to scale with body mass (Garland 1983); speed is also proportionate to body-length, a rule that applies across the range of running and swimming organisms from bacteria to arthropods to whales (Meyer-Vernet 2015), and as indicated by Jamieson et al. (2012), discussed in detail subsequently. Moreover, juveniles tend to be slower, weaker and less agile than adults of the same species (Carrier 1996). It is thus reasonable to assume that larger trilobites were capable of higher speeds than smaller trilobites. Also, because drafting generates reductions in metabolic and power requirements, it is reasonable to conclude that smaller trilobites could sustain speeds by drafting that were otherwise unsustainable when travelling in isolation. In this paper we model these effects.

Furthermore, we argue that published data of *T. chopini* size ranges and queue behaviors are consistent with the variation range hypothesis raised by Trenchard & Perc (2016). The variation range hypothesis posits that the size range among individuals in groups corresponds proportionately to the energy saving quantity (as a per cent) because weaker, smaller, individuals sustain speeds of stronger, larger, individuals by exploiting the energy saving mechanism. We determine size range by

$$SR = [(BL_{max} - BL_{min}) / BL_{max}] * 100, \qquad (1)$$

where *SR* is size range, and *BL* is body-length. As discussed further, this allows comparisons between size ranges and energy saving as a per cent, shown as equation (8).

Individuals too small to fit within this range become isolated from the group and may perish, or form sub-groups of narrower size ranges, as has been modelled by Trenchard et al. (2015) in the context of bicycle pelotons. In the wider context of migration as a factor that drives speciation (Winker 2000), we consider the possibility that this form of group-sorting may contribute to migratory divergence and to reproductive isolation, as proposed by Delmore et al. (2012).

Speyer & Brett (1985) reported that individuals within trilobite groups generally fall within a rather narrow size range, a finding which tends to support the variation range hypothesis. Moreover, the presence of any similarly sized trilobites of other species mixed with clusters of another species also tends to support the variation range hypothesis, since such individuals would likely have possessed similar power and speed capacities.

For example, Robison (1975) reported that several agnostid species (Wheeler formation, Utah) were aggregated within comparatively narrow size ranges in which inter-species' maximum pygidial length ratios were measured between 1.0 and 1.54. Hickerson (1997) reported mixed trilobite species associations in the Middle Devonian Little Cedar Formation (Illinois) (~44% mixed), in which clusters tended to be size-sorted. Some clusters included a specimen that was significantly smaller than the others and, in a few clusters, a size range was evident, although the specific size range was not reported (Hickerson 1997). Hickerson (1997) did report an instance of a 1.75 cm *Dechenella? prouti* on top of three ~3.5 cm *Phacops iowensis iowensis* (50% range), well within the range predicted by the variation range hypothesis. It is not known, however, whether these species engaged in queueing behavior or otherwise exploited hydrodynamic drafting during collective locomotion.

## METHODS

Trilobite queue formations may be understood as a relationship between the power output of the leading trilobite in the high-drag position, the energy reduction afforded by the energy saving mechanism, and the maximal power output of the following trilobite. Power output may be approximated in terms of speed assuming constant environmental conditions. Thus:

$$TCR = (S_{front} * D) / MSO_{follow}, \qquad (2)$$

where *TCR* is "trilobite convergence ratio"; $S_{front}$ is the speed of the trilobite in the high drag position, which sets the pace for followers; *D* is the drafting coefficient, or the proportion of power (here considered in terms of speed) required for a trilobite in a low drag (drafting) position relative to the power required for the trilobite in the leading, high drag, non-drafting position ($(1 - D)$ is the energy saving benefit to the following trilobite as a per cent); $MSO_{follow}$ is the maximal sustainable speed of following trilobites in reduced drag following positions. $MSO_{follow}$ varies according to the distance over which a given speed is sustained, and may also be reduced by fatigue.

Trilobites' sustainable relative speeds and their corresponding metabolic requirements are unknown. However, modern analogs provide some insights. For example, it is known that ghost crab endurance time as a function of speed drops rapidly as speeds increase and, in general, smaller animals attain maximum oxygen consumption at lower speeds than do larger animals (Full 1987). Similarly, it has been shown that when travelling up to 12 hours a day, terrestrial red crabs tend to travel at mean walking speeds near their maximum aerobic speed, using a mixed migration strategy in which pauses are interspersed with periods of locomotion that exceed crabs' maximum aerobic speed (Adamczeska & Morris 1998). By comparison, for human cyclists, the "power output profile" consists of five sustainable capacities: explosive, or "sprint" speeds, sustainable for ~10 seconds; lactic tolerance speeds, sustainable for ~30 s; maximal aerobic power, sustainable for ~5 min; anaerobic threshold, sustainable for between 20 and 60 min; and endurance, sustainable for up to several hours (Pinot & Grappe 2010).

With respect to queue formations, the durations for which trilobite queues were sustained are unknown. However, extant arthropods (spiny lobsters) are known to travel in queues for many hours (Bill & Herrnkind 1976), so we may consider trilobite *MSO* speeds that occurred over several minutes to hours before substantial output reductions were required. In long range migrations, we may expect short periods of "explosive" or lactic tolerance speeds during flight from predators, and relatively longer periods of endurance output. Or, like terrestrial red crabs, we may expect periods during which speeds exceeded maximum aerobic capacities, interspersed with pauses to rest or to feed. Moreover, if trilobites tended to migrate at mean speeds close to their maximum aerobic speeds as do terrestrial red crabs (Adamczeska & Morris 1998), then this would support the energy saving advantage of travelling in queues. Regardless, because data are not available to determine accurately the trilobite "power profile", we adopt a singular *MSO*~follow~ that corresponds to mid-range sustainable outputs and is constant for each trilobite individually in correspondence with their size, as discussed subsequently.

Our objective is thus to derive *T. chopini* travelling speeds and their maximal sustainable outputs ($MSO_{follow}$) and to evaluate whether observed size ranges for *T. chopini* within queues (Kin & Błażejowski 2013) are consistent with the variation range hypothesis.

Our analysis involves three primary steps. First, we demonstrate the availability of hydrodynamic drafting for trilobites. We determine a drafting coefficient (*D*) by applying the drag coefficient from Hoerner (1965) for a streamlined half-body, which approximates the lateral profile of *T. chopini*, as indicated in Fig. 2. Figure 2 shows the outstretched profile of a *Trimerocephalus interruptus*, a species of the same genus as *T. chopini*. This profile is consistent with inferred outstretched profiles of *T. chopini* shown in Błażejowski et al. (2015), and Kin & Błażejowski (2013).

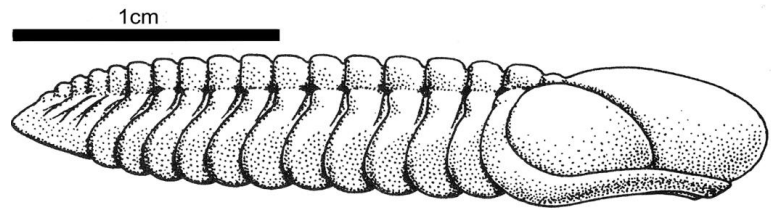

**Figure 2**. Outstretched *Trimerocephalus interruptus* (modified from Berkowski 1991, fig. 4B Creative Commons Attribution License (CC BY)), with streamlined half-body morphology like that of *T. chopini*.

Secondly, we determine realistic trilobite travelling and maximal sustainable speeds. For this we reconstruct speeds determined empirically for *Rectisura herculea,* a living crustacean of similar morphology to *T. chopini* and between ~2.5 and 5 times the size of *T. chopini* (Jamieson et al. 2012), to obtain a regression polynomial equation. We apply this equation to extrapolate travelling and maximal speeds of *R. herculea* for *T. chopini* body-length (*BL*) range 0.5 cm – 2.0 cm (75% range), as reported by Radwański et al. (2009). We note Błażejowski et al. (2016) reported a similar *BL* range of 0.7 cm to 1.9 cm (63% range) for *T. chopini* in queues.

Thirdly, we model trilobite collective behaviors, including the physical and physiological conditions underlying queuing behavior, and group sorting.

# RESULTS

*Hydrodynamic drafting availability for trilobites*

We have approximated drag forces for both the front trilobite and the drafting trilobite. Here "front trilobite" refers to a trilobite in the high-drag leading position, and "drafting trilobite" refers to a following trilobite that enjoys the energy saving of drafting. Drag force, $F_d$, is (Hoerner 1965):

$$F_d = ½ (C_d * p * S * V^2), \qquad (3)$$

where $C_d$ is the drag coefficient (dimensionless) and is equal to 0.07 for a streamlined half-body placed on a wall (Hoerner 1965, ch. 8, p. 8-4), characteristic of a benthic trilobite in locomotion on an ocean bed, and where the streamlined half-body shape is similar to *T. chopini* morphology, as shown in Fig. 2; $p$ is fluid density = 1027 kg m$^{-3}$ for seawater at 16°C; $S$ is frontal surface area (m$^2$) which we determined for a holaspid (late stage) *T. chopini* as 2.7 X 10$^{-5}$ m$^2$ by applying scaled length and height dimensions from Kin & Błażejowski (2013) (their fig. 4B) to a simplified cross-sectional elliptical shape; $V$ is relative velocity of fluid (m s$^{-1}$), equivalent to the velocity of the trilobite moving through water.

$C_d$ is largely a function of body shape, orientation, surface smoothness, and position relative to other bodies (Hoerner 1965; Denny 1989); flow velocity is less important (Chamberlain 1976). Hydrodynamic drag coefficient of a streamlined body approximately corresponds to aerodynamic drag coefficients (Hoerner 1965, ch. 10, p.10-4).

Drag coefficients consistent with those reported by Hoerner (1965) were reported for cephalopod shells with maximum diameter 12.7 cm (Chamberlain 1976) and are independent of body size (Hoerner 1965), while drag force scales with the size of the organism (Denny & Blanchette 2000).

We could not find reported drag coefficients for an in-wake streamlined half-body, but we may estimate that coefficient by reference to reported drag coefficients for two cylinders in tandem, where $C_d$ for the in-wake cylinder is 0.45 (Igarashi 1981), and $C_d$ for the cylinder in the leading high-drag position is 1.17 (Hoerner 1965, ch. 8; Igarashi 1981). The ratio of these two drag coefficients, 0.45/1.17, yields the drafting coefficient, $D$, 0.385 (dimensionless).

Thus, to approximate $C_d$ for the drafting half-streamlined body, we apply $D$=0.385, where 0.07 * 0.385 = 0.0269. Introducing $C_d$ = 0.0269 into equation (3) for the drafting trilobite, and $C_d$ = 0.07 for the front trilobite, we determine respective drag forces, $F_d$ (N), for the front half-streamlined body and the following half-streamlined body to approximate these values for trilobites over a range of travelling speeds up to 2.0 cm s$^{-1}$, as shown in Fig. 3.

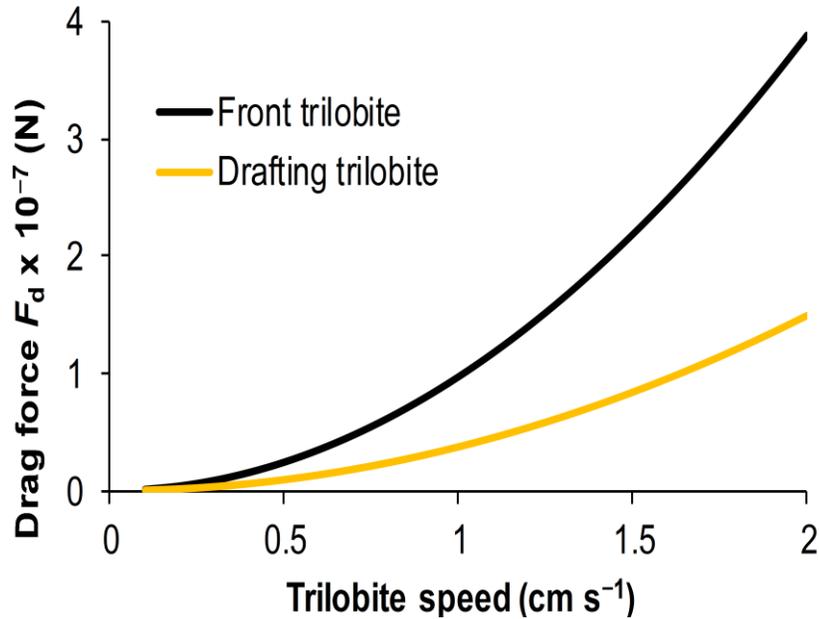

**Figure 3.** Drag forces (N) for trilobite in front, non-drafting position (upper curve), and for drafting trilobite (lower curve).

$D = 0.385$ may now be introduced into equation (2). The corresponding energy saving $(1 - D)$ is 0.615 (61.5%).

The approximated energy saving quantity of 61.5% is in reasonable agreement with Bill & Herrnkind's (1976) finding that spiny lobsters generate 65% energy savings by hydrodynamic drafting. It is also similar to the 63% maximum energy saving that Fish (1995) reported for ducklings swimming in a decoy wake compared to solitary swimming.

*Determining T. chopini travelling speeds and maximal sustainable speeds*

We used data from Jamieson et al. (2012), reconstructed in Fig. 4, for *R. herculea*, living crustaceans between 2.5 and 5 times the length of *T. chopini*, to derive an approximate range of speeds for *T. chopini* for a range of body-lengths (*BL*). To achieve this, for both sub-maximal and maximal walking speeds, we obtained a linear polynomial regression correlation between body-length and walking speeds for *R. herculea*, which permits us to extrapolate speeds for the shorter *T. chopini* body-lengths. We could find no data in the literature that more closely models arthropods of similar size, morphology and speeds to T. *chopini* by which to extrapolate approximate *T. chopini* speeds.

Jamieson et al. (2012) also reported a single individual *R. herculea* with a high size-specific speed of 0.33 cm s$^{-1}$ *BL*$^{-1}$ relative to the mean 0.19 cm s$^{-1}$ *BL*$^{-1}$. This particular *R. herculea* accelerated to 1.2 cm s$^{-1}$ as an aversion response in the presence of a fish, as shown in Fig. 4 (Jamieson et al. 2012). We consider this aversion response as within mid-range sustainable capacities described as a maximal aerobic power speed, sustainable for ~5 min; or an anaerobic threshold speed, sustainable for between 20 and 60 min (see foregoing discussion on the "power profile") for *R. herculea* (but not a maximally explosive speed, for which Jamieson et al. (2012) reported considerably higher backward jump escape speeds). From this aversion response speed, we approximated a scaled corresponding sustainable maximal aerobic power, or anaerobic threshold, speed for *T. chopini*.

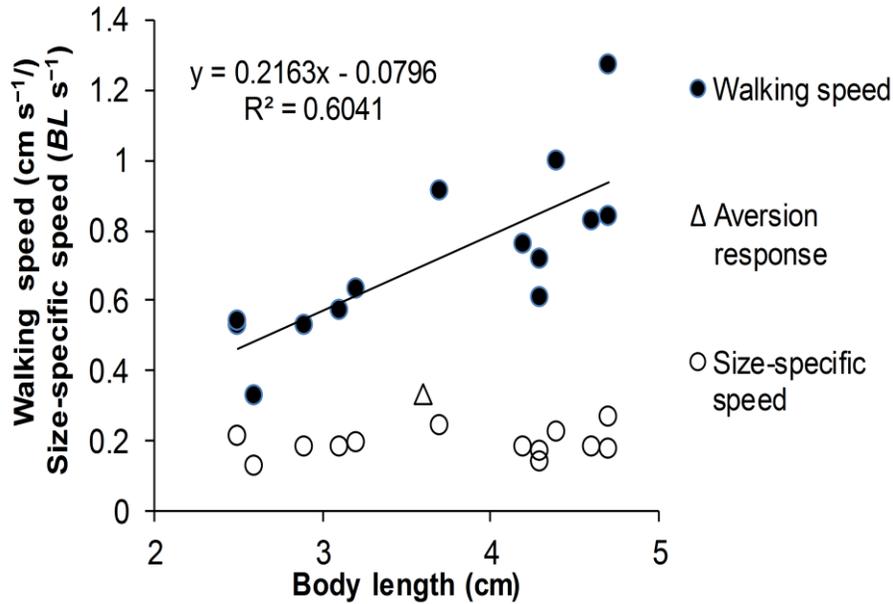

**Figure 4**. *R. herculea* absolute walking speeds (black circles) and size-specific speeds (open circles). Data reconstructed from Jamieson et al. (2012). Triangle denotes acceleration due to aversion response to fish nearby (corresponding maximum speed of 1.20 cm s$^{-1}$ not shown and discounted from regression polynomial to obtain best fit for normal, unperturbed, walking speeds). Data applied to approximate aerobic power, or anaerobic threshold, maximal travelling speeds, and sub-maximal *T. chopini* travelling speeds.

Applying regression polynomial y = 0.0216x – 0.0796, we extrapolated the following sub-maximal walking speeds for trilobites between 0.1 cm and 2.0 cm long, and aerobic power, or anaerobic threshold, maximal walking speeds based on a maximum size-specific speed of 0.33 *BL* s$^{-1}$, as shown in Table 1. To model coupling effects and thresholds, values from the third column are thus applied to establish realistic *MSO*$_{follow}$ and *S*$_{front}$ speeds, as shown in Figs 5 - 7.

| Body-length (cm) | [1]Estimated sub-maximal walking speed (cm s$^{-1}$) = 0.216 * *BL* - 0.0796 | [2]Estimated maximal sustainable walking speed (cm s$^{-1}$) |
|---|---|---|
| 0.1 | -0.058 | 0.033 |
| 0.2 | -0.0364 | 0.066 |
| 0.3 | [3]-0.0148 | 0.099 |
| 0.4 | 0.0068 | 0.132 |
| [4]**0.5** | **0.0284** | **0.165** |
| 0.6 | 0.05 | 0.198 |
| 0.7 | 0.0716 | 0.231 |
| 0.8 | 0.0932 | 0.264 |
| 0.9 | 0.1148 | 0.297 |
| 1 | 0.1364 | 0.33 |
| 1.1 | 0.158 | 0.363 |
| 1.2 | 0.1796 | 0.396 |
| 1.3 | 0.2012 | 0.429 |
| 1.4 | 0.2228 | 0.462 |
| 1.5 | 0.2444 | 0.495 |
| 1.6 | 0.266 | 0.528 |
| 1.7 | 0.2876 | 0.561 |
| 1.8 | 0.3092 | 0.594 |
| 1.9 | 0.3308 | 0.627 |
| **2** | **0.3524** | **0.66** |
| 2.1 | 0.374 | 0.693 |
| 2.2 | 0.3956 | 0.726 |
| 2.3 | 0.4172 | 0.759 |
| 2.4 | 0.4388 | 0.792 |
| 2.5 | 0.4604 | 0.825 |

[1] Estimated sub-maximal speeds for *T. chopini* of varying body-length.

[2] Estimated maximal sustainable walking speed trilobites of varying body-length, scaled according to aversion response size-specific speed of 0.33 cm s$^{-1}$ *BL*$^{-1}$ from Jamieson et al. (2012) for the extant crustacean, *R. herculea.*

[3] Negative values due to imprecision in regression equation and may be discounted for the model proposed.

[4] Max. and min. body-length values as reported by Kin & Błażejowski (2013) shown **bold.**

**Table 1.** Estimated *T. chopini* travelling speeds using regression polynomial equation obtained from data for *R. herculea* in Jamieson et al. (2012).

*Modelling group coupling and sorting behavior*

We can now model aggregate trilobite coupling and sorting behavior in terms of equation (2). We define trilobite coupling as the energetic or metabolic requirements of two trilobites whose outputs are determined by the speed set by the non-drafting trilobite, and the relative drag forces for the given position they occupy, whether non-drafting or drafting. Trilobites are optimally coupled when a following trilobite achieves maximum energy saving by hydrodynamic drafting. Varying degrees of coupling may occur depending on the distance between trilobites and the angle between leading and following positions, but trilobites remain coupled when drafting trilobites are *TCR* <= 1. If a weaker trilobite follows a stronger trilobite which sets a pace faster than the weaker trilobite's *MSO*, the weaker trilobite can sustain the pace of the stronger by drafting. De-coupling occurs when *TCR* > 1. Coupling between two trilobites extends globally to larger aggregates.

In Figs. 5 - 7 we show the effects of varying the three parameters: maximal speed capacity of the following trilobite ($MSO_{follow}$); speed set by the front trilobite ($S_{front}$); and the drafting coefficient, *D*, recalling that D = 0.385 and is the ratio of drag coefficients of an in-wake cylinder to a leading cylinder, 0.45 / 1.17, and the energy saving quantity is 1 – D = 0.615.

The greater the maximal capacity of a follower relative to the speed set by the leader, the longer the follower remains coupled to the leader. The smaller the energy saving value, the faster a stronger leader outpaces a follower; conversely, the greater the energy saving value, the longer the follower "sticks" with the leader.

Thus, it is apparent that if the $MSO_{follow}$ of a weaker trilobite exceeds the speed set by the front trilobite ($S_{front}$), as reduced by *D*, then the weaker trilobite can sustain the speed of a stronger trilobite by drafting. This is shown by the inequality,

$$MSO_{follow} \geq S_{front} * D. \qquad (4)$$

This condition is demonstrated in Fig. 5, where $MSO_{follow}$ is varied, $S_{front}$ is constant at 0.6 cm s$^{-1}$ and D = 0.385: all trilobites with $MSO_{follow} \geq$ 0.23 cm s$^{-1}$ are coupled with $S_{front}$ trilobites, and *TCR* < 1; all trilobites with $MSO_{follow}$ < 0.23 cm s$^{-1}$ are de-coupled from $S_{front}$ trilobites, and *TCR* > 1.

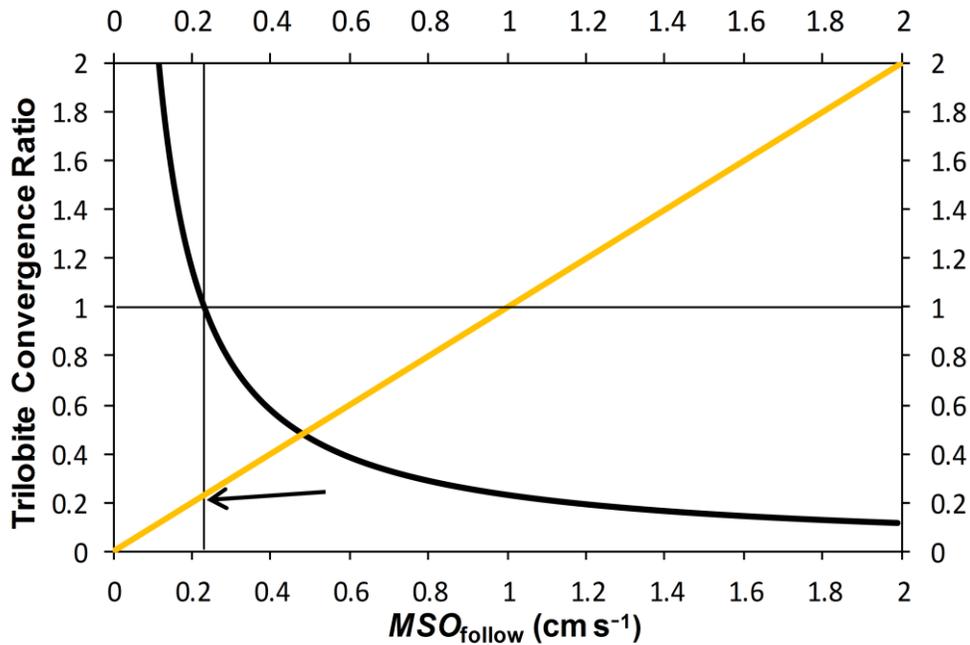

**Figure 5**. Coupling threshold for variable maximal sustainable output of drafting trilobite ($MSO_{follow}$). $MSO_{follow}$ is a variable between 0 and 2.0 cm s$^{-1}$ (diagonal curve). Trilobite convergence ratio = ($S_{front}$ * D) / $MSO_{follow}$ (non-linear black curve). $S_{front}$ is constant at 0.6 cm s$^{-1}$, and the drafting coefficient, $D$, is constant at 0.385 (energy saving quantity 1 – $D$ = 0.615). Arrow indicates threshold $MSO_{follow}$ (0.23 cm s$^{-1}$), when $TCR$ = 1. For all $MSO_{follow}$ > 0.23 cm s$^{-1}$ to right of vertical line when $TCR$ < 1, trilobites are coupled; for all $MSO_{follow}$ < 0.23 cm s$^{-1}$ to left of vertical line, when $TCR$ > 1, trilobites are de-coupled.

Similarly, if the speed of the front trilobite ($S_{front}$) does not exceed the maximum speed of the drafting trilobite as a ratio of the drafting coefficient, $D$, then it will not outdistance a drafting trilobite. This is shown by the inequality,

$$S_{front} \leq MSO_{follow} / D. \qquad (5)$$

This condition is demonstrated in Fig. 6, where $S_{front}$ is varied and $MSO_{follow}$ is constant at 0.3 cm s$^{-1}$ and $D$ = 0.385: all $S_{front}$ trilobites travelling ≤ 0.78 cm s$^{-1}$ are coupled with drafting trilobites, and $TCR$ < 1; all $S_{front}$ trilobites travelling > 0.78 cm s$^{-1}$ are de-coupled from drafting trilobites, and $TCR$ > 1.

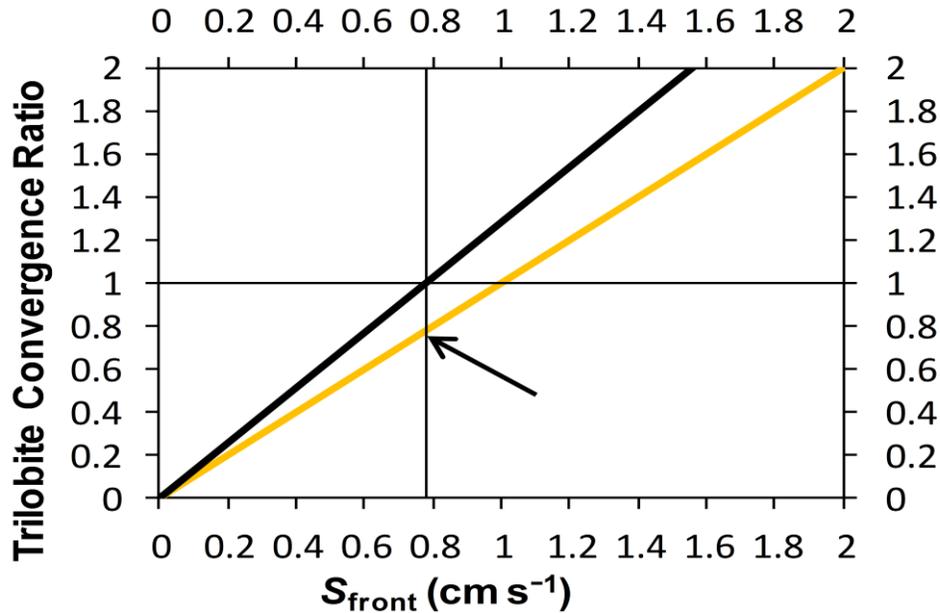

**Figure 6**. Coupling threshold for variable speed of leading trilobite ($S_{front}$). $S_{front}$ is a variable between 0 and 2.0 cm s$^{-1}$ (right diagonal curve). Trilobite convergence ratio = ($S_{front}$ * $D$) / $MSO_{follow}$ (left black diagonal curve). $MSO_{follow}$ is constant at 0.3 cm s$^{-1}$, and $D$ is constant at 0.385. Arrow indicates threshold speed of the leading trilobite ($S_{front}$ = 0.78 cm s$^{-1}$), where $TCR$ = 1. For all $S_{front}$ < 0.78 cm s$^{-1}$, left of vertical line when $TCR$ < 1, trilobites are coupled; for all $S_{front}$ > 0.78 cm s$^{-1}$ to the right of vertical line when $TCR$ > 1, trilobites are de-coupled.

If drafting coefficient, $D$, is less than the drafting trilobite's maximum speed as a proportion of the speed set by the front trilobite, then a drafting trilobite can sustain the pace of a stronger front trilobite. This is shown by the inequality,

$$D \leq MSO_{follow} / S_{front}. \tag{6}$$

This condition is demonstrated in Fig. 7, where $D$ is varied, $MSO_{follow}$ is constant at 0.3 cm s$^{-1}$ and $S_{front}$ is constant at 0.6 cm s$^{-1}$: for all $D \leq 0.501$, front and drafting trilobites are coupled, and $TCR$ < 1; for all $D > 0.501$, front and drafting trilobites are de-coupled, and $TCR$ > 1. The scenario in Fig. 7, where $D$ = 0.5, implies that a drafting trilobite, half the size of the front trilobite can, by exploiting the drafting benefit, sustain a pace set by the front trilobite that is twice the drafting trilobites' maximum speed when it travels alone. Here it should be recalled that as $D$ diminishes, the corresponding energy saving, 1 – $D$, increases.

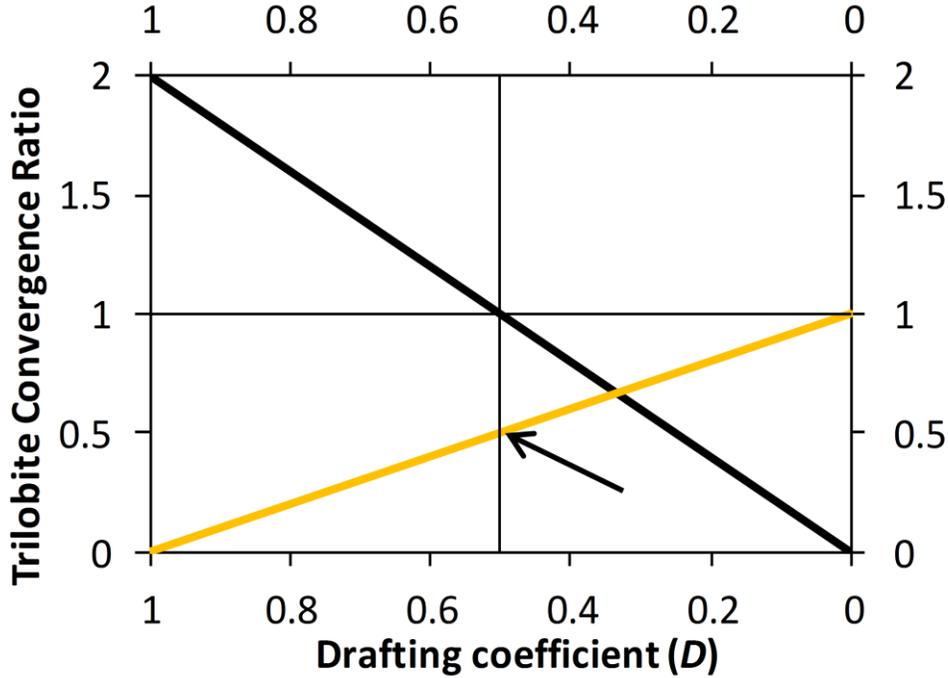

**Figure 7.** Coupling threshold for variable $D$. $D$ varies between 0 and 1.00 (lower diagonal curve); $S_{front}$ is constant at 0.6 cm s$^{-1}$; $MSO_{follow}$ is constant at 0.3 cm s$^{-1}$. Trilobite convergence ratio = $(S_{front} * D) / MSO_{follow}$ (upper black diagonal curve). Arrow indicates the threshold $D = 0.501$, when $TCR = 1$. For all $D < 0.501$ to the right of vertical line, when $TCR < 1$, all trilobites are coupled; for all $D > 0.501$, when $TCR > 1$, to the left of the vertical line, trilobites are de-coupled.

In turn, if the energy saving quantity, $1 - D$, exceeds the difference between the speed set by the front trilobite and the maximum speed of the follower, as a ratio of the speed of the front trilobite, the drafting trilobite will sustain the pace of the front trilobite. This is shown by the inequality,

$$1 - D \geq (S_{front} - MSO_{follow}) / S_{front}, \qquad (7)$$

where $S_{front} > MSO_{follow}$, and $1 - D = 0.615$.

The right-side term of equation (7) is in the same form as equation (1), which describes the size range ($SR$) for a set of trilobites. Since body-length corresponds to strength, as discussed, the basic inequality between the energy saving quantity and $SR$, foundational to the variation range hypothesis, is

$$1 - D \geq [(BL_{max} - BL_{min}) / BL_{max}]. \qquad (8)$$

Consequently, when the energy saving quantity, $1 - D$, exceeds the proportionately equivalent size range ($SR$) of a heterogeneous set of trilobites, the trilobites can remain coupled; conversely, trilobites outside $SR$ (i.e. too small) become decoupled.

# DISCUSSION

*Weaknesses in the hydrodynamic drafting model*

Although the energy saving quantity 61.5% is in good agreement with Bill & Herrnkind's (1976) finding that spiny lobsters generate 65% energy savings by hydrodynamic drafting, imprecision in our model exists due to the complexities of fluid dynamics. For instance, smaller trilobites will have had proportionately smaller frontal surface areas ($S$) than their larger counterparts. $S$ is typically calculated using two different methods: first, total frontal surface area given a convoluted surface morphology ("wetted area"); second, the maximum cross-sectional area; each yield significantly different drag forces and it may be recommended that both methods be used for accuracy (Alexander 1990).

Our simplified model is based on cross-sectional frontal surface areas for leading and following trilobites that are equal, but where trilobite body-length is varied. Since body-length is the critical parameter that correlates to an organism's speed (Meyer-Vernet 2015), this permits us to discount frontal surface area in our simplified analysis.

Furthermore, the drafting coefficient $D = 0.385$ does not account for a region of negative drag (a "suction zone") reported for a critical spacing up to 3.5 diameters between cylinders (Igarashi 1981). However, trilobite morphology is not cylindrical; rather it is an elongated streamlined profile that fills much of the suction zone that exists between cylinders. Therefore, we do not model the effect of the streamlined trilobite body on the magnitude of this suction zone.

A further consideration is the change in drag interaction between cylinders with respect to orientation angle (Dalton & Szabo 1977). Thus, for trilobites in moving queues, directional changes laterally and vertically will have had large effects on their drag forces and energetic demands. Moreover, the optimal drafting orientation was likely offset in the order of 30° laterally to the midline of the front trilobite; this is indicated for cylinders in tandem in which a free-floating drafting cylinder can overtake the one ahead (Tchieu et al. 2010), thereby also avoiding collision. Similar offset optimal drafting orientations are observed among fish species, although for different reasons (Marras et al. 2015).

A refined hydrodynamic model should also consider the costs of trilobite collisions and collision-avoidance, both of which appear to have occurred frequently, as shown in Fig. 1D and in the fossil record as shown in Radwański et al. (2009).

In general, the fluid dynamics that may be considered for a refined hydrodynamic model of trilobites in motion relative to each other involve complex equations of motion and fluid turbulence (Tchieu et al. 2010; Borisov et al. 2007) which are beyond the scope of this paper. Future studies may account for these fluid dynamical complexities, and may also involve computer simulations or the use of physical "toy" model trilobites to measure drag forces and optimal drag reducing positions, like the experiments of Bill & Herrnkind (1976) for spiny lobsters.

*The "trilobite convergence ratio" and implications for trilobite collective behavior*

The "trilobite convergence ratio" (*TCR*) model allows several inferences about the collective behavior of trilobites. As discussed, as trilobite speeds increase to approach their *MSO*s, the group may be expected to align into queues, as observed to occur among cyclists (Trenchard et al. 2014) and shown in Fig. 1A. Because of the hydrodynamic drafting coefficient that permits trilobite energy saving up to ~62%, we may expect correspondingly weaker trilobites to sustain the pace of stronger trilobites by exploiting drafting positions.

It should be noted that while cyclists gain substantial energy reduction advantages by remaining in pelotons, theirs is ultimately a competition in which the objective is to finish ahead of other cyclists. This is often achieved by intentionally seeking to separate oneself ahead of the peloton. For trilobites, particularly blind forms such as *Trimerocephalus*, we may expect some advantage conferred by separating ahead of the aggregation for faster access to foraging resources, but this may have been outweighed by the cost of becoming permanently isolated from the group. Consequently, particularly for blind trilobites, the survival advantage of remaining in close contact may have outweighed any advantages to advancing ahead of the group. Also, as with modern spiny lobsters, trilobites may well have travelled with antennae in contact with the individual ahead to keep together for defensive, feeding and/or other functions. Antennae were revealed in T. *chopini* by X-ray microcomputed tomography (Kin and Błażejowski 2013), and the distances indicated in the fossil record are generally sufficiently short to permit antennae contact (Radwański et al. 2009, pl. 1-5).

Radwański et al. (2009) observed that longer single-file trilobite lines were primarily formed of larger individuals, while smaller individuals were observed to comprise shorter queues, generally of two specimens. Although not without other explanations, in this case the fossil record may have captured the sorting process in action in which smaller trilobites were in the process of being isolated from larger ones. In circumstances of group separation like this, if the larger trilobites maintained speeds that established a separation between themselves and weaker trilobites, eventually the distance between themselves and the smaller ones may have been too great for the groups to reintegrate.

Despite being able to sustain the speeds of larger trilobites by exploiting hydrodynamic drafting, small following trilobites may be expected to have spent more time at near maximal outputs even in drafting positions. Larger trilobites, weakened during periods spent in high-drag positions, would be expected to be overtaken by other large trilobites, thus sharing the high-drag positions and maintaining a constant high collective speed while overtaken large trilobites would recuperate in drafting positions. This high-drag position sharing is a function of natural decelerations due to leader fatigue. The fatigued individuals are then passed by "fresh" trilobites. This kind of self-organized passing behavior is not due to any volitional motivation to share positions. Bill & Herrnkind

(1976) observed similar positional sharing among spiny lobsters, although the authors did not quantify or explain the behavior. Similar positional sharing has been observed among spiderlings travelling in single file, although the reasons for this behavior have not been proposed (Reichling 2000). Positional sharing among cyclists in pelotons due to fatigue is well known (Olds 1998).

At corresponding low speeds, small trilobites may have been capable of overtaking weakening larger trilobites to assume a high drag position, but only with a drastic reduction in group speed. In these circumstances, larger trilobites would be expected to quickly overtake the smaller ones, returning the group to a higher speed. We may therefore expect that larger trilobites generally tended to set the aggregate pace.

*Trilobite de-coupling and sub-group formation.* In circumstances in which larger trilobites tended to set the aggregate pace, smaller trilobites likely weakened faster than larger ones even while exploiting drafting positions. In turn, eventually smaller trilobites would weaken to the critical threshold condition $TCR > 1$ and de-couple from the group of larger trilobites. Moreover, de-coupling may have occurred even when stronger trilobites accelerated the pace at their own sub-maximal speeds.

This greater fatigue rate among weaker trilobites could occur even while generally enjoying drafting advantages because weaker trilobites were likely to be highly susceptible to positional instabilities, even at intermediate speeds set by stronger front trilobites. If caught out of optimal drafting range or position, even momentarily, small trilobites would have been required to increase their output dramatically to sustain the speeds set by the leaders or to catch back up and relocate to optimal drafting positions. Such a process would have been highly fatiguing, eventually producing the condition $TCR > 1$.

This describes the localized sorting process. Extended globally over the course of a migration period, we infer that groups underwent a continuous sorting process leading to the formation of sub-groups in which sub-groups tended to consist of members of similar strength, and corresponding size. This strength sorting principle was modelled by Trenchard et al. (2015) in the context of groups of cyclists. Trenchard & Perc (2016) proposed that this sorting process also depends partly on the number of group members, since the larger the group, the longer the sorting process.

*Size ranges greater than the hypothesized variation range may indicate pre-migratory stages.* Because the hypothesized sorting process probably occurred over a comparatively long time-span, trilobite aggregations whose size ranges exceeded the hypothesized variation range (equivalent to $1 - D$, or ~62%) were likely in pre- or early migratory stages when the process of sorting by size was not yet underway or complete. Early instars probably did not participate in migrations, but may have been involved in short–range foraging commutes.

Such broad size ranges most likely occurred if adults were mixed with much smaller (i.e. smaller than within the hypothesized ~62% range) early stage instars after hatching, such as in the case where adults co-existed with juvenile "nurseries", as suggested by Błażejowski et al. (2016).

A wide size range of this kind was reported by Lin & Yuan (2009), who studied a *Pagetia* trilobite cluster of 22 individuals from the Middle Cambrian Kaili Formation Guizhou (China). The cluster contained both degree 0 meraspids with cephalic length as short as 0.44 mm, and holaspids with cephalic lengths and long as 2.17 mm (80% range).

Similarly, Cederström et al. (2011) reported meraspid and holaspid individuals of the early Cambrian *Strenuaeva inflata* in the Tornesträsk area (Sweden) across a wide size range: for 3397 specimens, cranidia measurements ranged between ~ 0.2 cm and 2.0 cm (90% range). The authors argued that the aggregations occurred for brooding purposes and that the absence of very small instars could be explained by earliest-stage development in brood pouches (Cederström et al. 2011). If, in fact, the reported aggregations were in a brooding and early developmental stage, this implies a pre-migratory stage with respect to early instars which would not yet have undergone migration (but post-migration for adults having arrived prior to brooding).

*Size range less than or approximately equal to the hypothesized variation range indicate stopovers or post-migratory stages.* Generally, the argument we present supports the proposition that queues occurred during migration (Radwański et al. 2009; Błażejowski et al. 2016), while non-linear clusters formed during sedentary "stopover" periods, when moulting, brooding, or spawning.

Although non-linear clusters were probably not in a travelling phase, owing to their random orientations (Karim & Westrop 2002; Speyer and Brett 1985), it is reasonable to expect the trilobites migrated initially before arriving at these clustering locations. It was during migratory periods when size sorting would occur, and so we would expect to see evidence among the clusters during "stopovers" of size-sorting processes that occurred during migration periods. Narrow size ranges observed among spawning, brooding, or moulting clusters is just the sort of evidence that tends to support the variation range hypothesis.

Speyer & Brett (1985) reported size-segregated clusters of Middle Devonian *Phacops rana*, *Greenops boothi,* and *Dechenella rowi* from the Windom Smoke Creek Bed (Windom Shale, Hamilton Group), and from the Murder Creek Bed (Wanakah Shale) both from western New York State. The authors reported cephalon length ranges of 0.6 to 1.4 cm (57% range, where cephalon length correlates with body-length (Trammer & Kaim 1997)), and cephalon length ranges of 0.4 to 1.0 cm (60% range). Further, the authors reported spatially separated clusters of different mean size, indicating that specific instar classes associated among themselves to the exclusion of other classes.

In a similar finding, Karim & Westrop (2002) reported a non-linear cluster of Late Ordovician *Homotelus bromidensis* from the Bromide Formation, Dunn Quarry (Oklahoma) with cephalic lengths between 1.0 cm and 2.25 cm (56% range), and a second non-linear cluster with cephalic lengths between 1.0 cm and 2.75 cm (64% range).

These cases indicate group members travelled together due to their approximate size equality, and suggest that groups of different mean speeds would arrive at stopover points at different times. This proposition does not challenge a gregarious behavioral explanation for instar segregation, but rather complements such an explanation while providing insight into the more primitive origins of gregarious segregation.

Kin & Błażejowski (2013) reported that among 78 examples of Late Devonian (Famennian) *Trimerocephalus* queues from the Kowala Quarry (Poland), specimens ranged in size from 0.5 cm – 2.0 cm body-length (75% range). Although this range exceeds the range of ~62% predicted by the variation range hypothesis, the overall 0.5 cm – 2.0 cm body-length (75% range) appears to represent the size range among all specimens in the study but does not distinguish between size segregated groups and narrower size-ranges among the queues themselves.

Following the Kin & Błażejowski (2013) study, Błażejowski et al. (2016) reported that for the same 78 queues, the size ranges for individual queues were between 0.7 and 1.9 cm (63% range), thus supporting the assertion that the 75% range reported by Kin & Błażejowski (2013) was for the entire sample population and not queue-specific.

It is also noteworthy that in their study of queues from the Kowala Quarry, Radwański et al. (2009) reported that "The majority of queues are formed from the largest individuals. The smaller-sized individuals are arranged as a rule in short files consisting of only two individuals" (p. 467). From this it appears that the queues had indeed sorted themselves in much the way predicted by the variation range hypothesis. Kin & Radwański (2008) also reported specimens from the Kowala Quarry in files, of mature growth stage, between 1.8 and 2.4 cm (25% range).

We note that enrolled juvenile *T. chopini* smaller than within this range were reported by Błażejowski et al. (2016), situated a few millimetres above the queue formation bedding plane. However, the nearby presence in the fossil record of smaller enrolled juveniles presents a challenge to the variation range hypothesis in these circumstances, because it suggests queues were in travelling mode somewhat concurrently with the presence of nearby small juveniles which, according to the variation range hypothesis, would have been isolated and left behind by the travelling queues.

However, because they appear on a different bedding plane, the small enrolled juveniles may have arrived at their positions at a different time, or may have already been at their positions in "nursery grounds" before the queues arrived as Błażejowski et al. (2016) tentatively explained, and probably did not migrate with the queues containing their much larger counterparts.

In another study, Gutiérrez-Marco et al. (2009) reported monospecific clusters of large Middle Ordovician trilobites *Ogyginus forteyi*, and *Asaphellus* in Arouca Geopark (Portugal), 7 to 17 in number, were "similar sized specimens" (p.444), but the authors did not report precise ranges.

The energy reduction hypothesis for trilobite queueing behavior does not exclude single-file formations observed in the fossil record for other reasons, such as trilobites having been constrained in linear burrows (Chatterton & Fortey 2008). However, it may be an alternative explanation to proposed chemosensory coordination for blind trilobites (Błażejowski et al. 2016). By exploiting energy saving positions behind leading trilobites, followers may have modified their positions according to sustainable metabolic outputs and naturally gravitated toward the "easiest" positions. This principle of local self-organized behavior would lead to queue formation at the aggregate level, as observed to occur in bicycle pelotons (Trenchard et al. 2015). Moreover, because collective behavior can emerge among inanimate objects in the presence of hydrodynamic drafting (Wang et al. 2015), it is possible the energy saving mechanism for queues preceded the development of trilobite chemosensory apparatus' in the evolutionary lineage.

**CONCLUSIONS**

We have proposed that by travelling in queues, trilobites could exploit the energy saving benefits of hydrodynamic drafting. Weaker trilobites could thereby sustain speeds, as set by stronger trilobites, which were otherwise unsustainable when travelling in solitary isolation. We have proposed that the degree to which trilobites could have been weaker (and therefore smaller in size) and still have sustained speeds of stronger trilobites, approximately corresponds with the degree of energy saved by drafting.

We have modelled certain parameters that define trilobite coupling and the sorting process, including the relationship between the pace set by the leading trilobite, the maximal sustainable output of the following trilobite, and the energy saving mechanism of hydrodynamic drafting.

We have proposed that the correspondence between trilobite strength (and size) and the energy saved by drafting, implies a process by which trilobite groups tended to sort into sub-groups of increasingly narrow size ranges. Consequently, during migration the maximum size range we would expect to observe is ~62%, since this is approximately the quantity of energy saved by hydrodynamic drafting. Over extended migrations, we would expect this range to narrow increasingly because of a hypothesized faster fatigue rate for weaker trilobites, even with the availability of drafting, due to positional instabilities and their temporary exposure to high-drag positions. Conversely, we suggest that stronger trilobites, by self-organized continuous collective motion, may have shared drafting positions and distributed the time spent in high-drag positions among group members, thereby resulting in a net advantage for stronger trilobites over time. Consequently, the longer the migration, the more likely it was for trilobites to sort into groups of narrow size ranges.

Trilobites had to be adaptive to the various selection pressures at all growth stages because, of course, they could not reach adult sizes without first going through their instar stages and corresponding sizes. However, the proposed model, by which groups sort into narrow size ranges and by which those outside the threshold range may be de-coupled from the group and become reproductively isolated, is not in conflict with trilobite fitness at all instar stages. This is because de-coupling and size sorting would occur only during locomotion and migration periods of sufficient duration, and this process would be scaled in correspondence to the different mean sizes of each aggregation travelling independently.

It is likely that individuals among early instar stages up to a certain size did not participate in queues simply because they were not sufficiently mature to migrate. However, this raises the question as to why members of each instar stage would proceed through their growth stages within certain size ranges in the first place. The proposed model addresses this: because spawning or brooding trilobites, arriving after the proposed migratory sorting process had occurred, would naturally produce hatchlings whose sizes were consistent with the size range of those trilobites which survived the journey to spawn or brood. This suggests, therefore, that hatchlings would generally begin life within this approximate size range (as a per cent) and commence their own migrations as a group (when sufficiently strong to do so), scaled in sizes according to their instar stage and growing accordingly with the passage of time, but travelling separately from their larger adult counterparts. Thereupon the process would be repeated.

Alternatively, and perhaps concurrently, trilobites of varying sizes may have had different behaviors, and that queue formation was only a behavior typical of individuals above a given size. This would be in line with variations in trilobite size ranges observed in different environments. In any case, the examples of fossilized trilobite queuing behavior are presently comparatively few, and more data is required to draw inferences with any degree of certainty. Fossil records in which trilobites were travelling unidirectionally, not in queues but in more compact formations, may also assist our understanding of the advantages of queue formations, in hydrodynamic terms or otherwise.

Future work may involve modelling more precisely the complex fluid dynamics involved in trilobite hydrodynamic drafting and its energetic consequences, including "toy" model experiments. In addition, future work may include a more comprehensive review of the reported size ranges among fossilized trilobites and other species and an analysis of their corresponding migratory stages. The proposed model presents a novel paleoecological framework in which to consider migration and size variation among trilobites, in addition to other fossil and extant organisms.

**REFERENCES**

ADAMCZEWSKA, A. and MORRIS, S. 1998. Strategies for migration in the terrestrial Christmas Island red crab *Gecarcoidea natalis*: intermittent versus continuous locomotion. *Journal of experimental biology*, **201** (23), 3221-3231.

ALEXANDER, D.E. 1990. Drag coefficients of swimming animals: effects of using different reference areas. *The Biological Bulletin*, **179** (2), 186-190.

BERKOWSKI, B. 1991. A blind phacopid trilobite from the Famennian of the Holy Cross Mountains. *Acta Palaeontologica Polonica*, **36** (3), 255-264.

BILL, R.G., and HERRNKIND, W.F. 1976. Drag reduction by formation movement in spiny lobsters. *Science*, **193**, 1146-1148.

BŁAŻEJOWSKI, B., BRETT, C.E., KIN, A., RADWAŃSKI, A. and GRUSZCZYŃSKI, M. 2016. Ancient animal migration: a case study of eyeless, dimorphic Devonian trilobites from Poland. *Palaeontology*, **59** (5), 743-751.

BŁAŻEJOWSKI, B., GIESZCZ, P., BRETT, C.E. and BINKOWSKI, M., 2015. A moment from before 365 Ma frozen in time and space. *Scientific Reports*, **5**, Article 14191.

BORISOV, A.V., MAMAEV, I.S. and RAMODANOV, S.M. 2007. Dynamics of two interacting circular cylinders in perfect fluid. *Discrete and Continuous Dynamical Systems*, **19** (2), 235-253

BRADDY, S.J., 2001. Eurypterid palaeoecology: palaeobiological, ichnological and comparative evidence for a 'mass–moult–mate' hypothesis. *Palaeogeography, Palaeoclimatology, Palaeoecology*, **172** (1), 115-132.

CARRIER, D.R. 1996. Ontogenetic limits on locomotor performance. *Physiological Zoology*, **69** (3), 467-488.

CEDERSTRÖM, P., AHLBERG, P., NILSSON, C.H., AHLGREN, J. and ERIKSSON, M.E. 2011. Moulting, ontogeny and sexual dimorphism in the Cambrian ptychopariid trilobite *Strenuaeva inflata* from the northern Swedish Caledonides. *Palaeontology*, **54** (3), 685-703.

CHAUDHURI, D. and NAGAR, A. 2015. Absence of jamming in ant trails: Feedback control of self-propulsion and noise. *Physical Review E*, **91** (1), 1-5.

CHAMBERLAIN, J.A. 1976. Flow patterns and drag coefficients of cephalopod shells. *Palaeontology*, **19** (3), 539-563.

CHATTERTON, B.D.E. and FORTEY, R.A. 2008. Linear clusters of articulated trilobites from Lower Ordovician (Arenig) strata at Bini Tinzoulin, north of Zagora, southern Morocco. *Advances in trilobite research. Cuadernos del Museo Geominero*, **9**, 73-77.

DALTON, C. and SZABO, J. 1977. Drag on a group of cylinders. *Journal of Pressure Vessel Technology*, **99** (1), 152-157.

DELMORE, K.E., FOX, J.W. and IRWIN, D.E. 2012. Dramatic intraspecific differences in migratory routes, stopover sites and wintering areas, revealed using light-level geolocators. *Proceedings of the Royal Society of London B: Biological Sciences*, **279** (1747), 4582-4589.

DENNY, M. 1989. A limpet shell shape that reduces drag: laboratory demonstration of a hydrodynamic mechanism and an exploration of its effectiveness in nature. *Canadian Journal of Zoology*, **67** (9), 2098-2106.

DENNY, M.W. and BLANCHETTE, C.A. 2000. Hydrodynamics, shell shape, behavior and survivorship in the owl limpet *Lottia gigantea*. *Journal of Experimental Biology*, **203** (17), 2623-2639.

DINGLE, H. and DRAKE, V.A., 2007. What is migration? *Bioscience*, **57** (2), 113-121.

DRAGANITS, E., GRASEMANN, B. and BRADDY, S.J. 1998. Discovery of abundant arthropod trackways in the? Lower Devonian Muth Quartzite (Spiti, India): implications for the depositional environment. *Journal of Asian Earth Sciences*, **16** (2), 109-118.

FISH, F.E. 1995. Kinematics of ducklings swimming in formation: consequences of position. *Journal of Experimental Zoology Part A: Ecological Genetics and Physiology*, **273** (1), 1-11.

FITZGERALD, T.D., PESCADOR-RUBIO, A., TURNA, M.T. and COSTA, J.T. 2004. Trail marking and processionary behavior of the larvae of the weevil *Phelypera distigma* (*Coleoptera: Curculionidae*). *Journal of Insect Behavior*, **17** (5), 627-646.

FULL, R.J., 1987. Locomotion energetics of the ghost crab: I. Metabolic cost and endurance. *Journal of Experimental Biology*, **130** (1), 137-153.

GAILLARD, C., HANTZPERGUE, P., VANNIER, J., MARGERARD, A-L., and MAZIN, J-M. 2005. Isopod trackways from the Crayssac Lagerstätte, Upper Jurassic, France. *Paleontology.* **48** (5), 947-962.

GARLAND, T. JR. 1983. The relation between maximal running speed and body mass in terrestrial mammals. *Journal of Zoology.* **199** (2), 157-170.

GUTIÉRREZ-MARCO, J.C., SÁ, A.A., GARCÍA-BELLIDO, D.C., RÁBANO, I. and VALÉRIO, M. 2009. Giant trilobites and trilobite clusters from the Ordovician of Portugal. *Geology*, **37** (5), 443-446.

HANSEN, L.D. AND KLOTZ, J.H. 2005. *Carpenter ants of the United States and Canada.*Cornell University Press. New York, 204 pp.

HEINRICH, B., VOGT, F.D. 1980. Aggregation and foraging behavior of whirligig beetles (Gyrinidae). *Behavioral Ecology and Sociobiology*, **7** (3), 179-186.

HERRNKIND, W. F., KANCIRUK, P., HALUSKY, J. and MCLEAN, R. 1973. Descriptive characterization of mass autumnal migrations of spiny lobster, *Panulirus argus. Proceedings of the Gulf and Caribbean Fisheries Institute,* **25**, 78–98.

HICKERSON, W.J. 1997. Middle Devonian (Givetian) trilobite clusters from eastern Iowa and northwestern Illinois. *Paleontological Events: Stratigraphic, Ecological, and Evolutionary Implications. Columbia University Press, New York*, 224-246 (eds. Brett, C.E. & Baird, G.C).

HOERNER, S.F. 1965. *Fluid-dynamic drag: practical information on aerodynamic drag and hydrodynamic resistance*. Hoerner Fluid Dynamics, New Jersey, 455 pp.

HOU, X.G., SIVETER, D.J., ALDRIDGE, R.J. and SIVETER, D.J. 2008. Collective behavior in an Early Cambrian arthropod. *Science*, **322** (5899), 224-224.

IGARASHI, T. 1981. Characteristics of the flow around two circular cylinders arranged in tandem: 1st report. *Bulletin of the Japan Society of Mechanical Engineers*, **24** (188), 323-331.

JAMIESON, A.J., FUJII, T. and PRIEDE, I.G. 2012. Locomotory activity and feeding strategy of the hadal munnopsid isopod *Rectisura* cf. *herculea* (Crustacea: Asellota) in the Japan Trench. *Journal of Experimental Biology*, **215** (17), 3010-3017.

KARIM, T. and WESTROP, S.R. 2002. Taphonomy and paleoecology of Ordovician trilobite clusters, Bromide Formation, south-central Oklahoma. *Palaios*, **17** (4), 394-402.

KIN, A. and BLAŻEJOWSKI, B. 2013. A new *Trimerocephalus* species (Trilobita, Phacopidae) from the Late Devonian (Early Famennian) of Poland. *Zootaxa*, **3626** (3), 345-355.

KIN, A. and RADWAŃSKI, A. 2008. Trimerocephalus life-position files from the Famennian of Kowala Quarry (Holy Cross Mountains, Central Poland). Polish Academy of Sciences, Institute of Paleobiology. In *9th Paleontological Conference, Warszawa* 10-11 October 2008; Abstracts, 44.

MARRAS, S., KILLEN, S.S., LINDSTRÖM, J., MCKENZIE, D.J., STEFFENSEN, J.F. and DOMENICI, P. 2015. Fish swimming in schools save energy regardless of their spatial position. *Behavioral Ecology and Sociobiology*, **69** (2), 219-226.

LIEDVOGEL, M., ÅKESSON, S., and BENSCH, S. 2011. The genetics of migration on the move. *Trends in Ecology and Evolution*, **26** (11), 561-569.

LIN, J.P. and YUAN, J.L., 2009. Reassessment of the mode of life of *Pagetia* Walcott, 1916 (*Trilobita*: *Eodiscidae*) based on a cluster of intact exuviae from the Kaili Formation (Cambrian) of Guizhou, China. *Lethaia*, **42** (1), 67-73.

MEYER-VERNET, N. 2015. How fast do living organisms move: Maximum speeds from bacteria to elephants to whales. *American Journal of Physics*, **83**, 718-722.

OLDS, T. 1998. The mathematics of breaking away and chasing in cycling. *European Journal of Applied Physiology and Occupational Physiology*, **77** (6), 492-497.

PINOT, J. and GRAPPE, F. 2011. The record power profile to assess performance in elite cyclists. *International Journal of Sports Medicine*, **32** (11), 839-844.

RADWAŃSKI, A., KIN, A. and RADWANSKA, U. 2009. Queues of blind phacopid trilobites *Trimerocephalus*: A case of frozen behaviour of Early Famennian age from the Holy Cross Mountains, Central Poland. *Acta Geologica Polonica*, **59** (4), 459-481.

REICHLING, S.B., 2000. Group dispersal in juvenile *Brachypelma vagans* (*Araneae, Theraphosidae*). *Journal of Arachnology*, **28** (2), 248-250.

RITZ, D.A., 2000. Is social aggregation in aquatic crustaceans a strategy to conserve energy? *Canadian Journal of Fisheries and Aquatic Sciences*, **57** (S3), 59-67.

ROBISON, R. 1975. Species diversity among agnostoid trilobites. *Fossils and Strata*, **4**, 219-226.

SPEYER, S.E. and BRETT, C.E. 1985. Clustered trilobite assemblages in the Middle Devonian Hamilton group. *Lethaia*, **18** (2), 85-103.

STEINBAUER, M.J., 2009. Thigmotaxis maintains processions of late-instar caterpillars of Ochrogaster lunifer. *Physiological Entomology*, **34** (4), 345-349.

TRAMMER, J. and KAIM, A. 1997. Body size and diversity exemplified by three trilobite clades. *Acta Palaeontologica Polonica*, **42** (1), 1-12.

TRENCHARD, H., 2015. The peloton superorganism and protocooperative behavior. *Applied Mathematics and Computation*, **270**, 179-192.

TRENCHARD, H. and PERC, M. 2016. Energy saving mechanisms, collective behavior and the variation range hypothesis in biological systems: A review. *Biosystems*, **147**, 40-66.

TRENCHARD, H., RICHARDSON, A., RATAMERO, E. and PERC, M., 2014. Collective behavior and the identification of phases in bicycle pelotons. *Physica A: Statistical Mechanics and its Applications*, **405**, 92-103.

TRENCHARD, H., RATAMERO, E., RICHARDSON, A. and PERC, M., 2015. A deceleration model for bicycle peloton dynamics and group sorting. *Applied Mathematics and Computation*, **251**, 24-34.

TCHIEU, A.A., CROWDY, D. and LEONARD, A. 2010. Fluid-structure interaction of two bodies in an inviscid fluid. *Physics of Fluids*, **22** (10), 107101-1-12.

WANG, L., GUO, Z.L. and Mi., J.C. 2014. Drafting, kissing and tumbling process of two particles with different sizes. *Computers & Fluids*, **96**, 20-34.

WILSON, E. 1959. Communication by tandem running in the ant genus *Cardiocondyla*. *Psyche*, **66** (3), 29-34.

WINKER, K. 2000. Migration and speciation. *Nature*, **404**, 36.

XIAN-GUANG, H.O.U., SIVETER, D.J., ALDRIDGE, R.J. and SIVETER, D.J. 2009. A new arthropod in chain-like associations from the Chengjiang Lagerstätte (Lower Cambrian), Yunnan, China. *Palaeontology*, **52** (4), 951-961.

YEN, J., BROWN, J. and WEBSTER, D.R. 2003. Analysis of the flow field of the krill, *Euphausia pacifica*. *Marine and Freshwater Behaviour and Physiology*, **36** (4), 307-319.